\documentclass[twocolumn]{svjour3}   
\smartqed  
\usepackage{tabularx} 
\usepackage{amsmath}  
\usepackage{amssymb}
\usepackage{bm}
\usepackage{caption}
\usepackage{float}
\usepackage{graphicx} 
\usepackage{gensymb}
\usepackage[numbers]{natbib}
\usepackage[final]{hyperref} 
\usepackage{color}

\journalname{Granular Matter}
\begin{document}

\title{Dynamics of oblique impact in a photoelastic granular medium}

\author{Cacey Stevens Bester $^{1,2}$ \and
        Noah Cox $^{2}$ \and
        Hu Zheng $^{2,3,4}$ \and
        Robert P. Behringer $^{2}$}

\institute{
            C. S. Bester\\
            \email{cbester1@swarthmore.edu}\\
            $^1$ Department of Physics and Astronomy, Swarthmore College, Swarthmore, PA, 19081, USA \\
            $^2$ Department of Physics \& Center for Non-linear and Complex Systems, Duke University, Durham, North Carolina, 27708, USA\\
            $^3$ Department of Geotechnical Engineering, College of Civil Engineering, Tongji University, Shanghai, 200092, China\\
            $^4$ School of Earth Science and Engineering, Hohai University, Nanjing, Jiangsu, 211100, China\\
}

\date{Received: date / Accepted: date}

\maketitle

\begin{abstract}
 When a solid projectile impacts a granular target, it experiences a drag force and abruptly comes to rest as its momentum transfers to the grains. 
 An empirical drag force law successfully describes the force experienced by the projectile, and the corresponding grain-scale mechanisms have been deciphered for normal impacts.  
 However, there is little work exploring non- normal impacts.  
 Accordingly, we extend studies to explore oblique impact, in which a significant horizontal component of the drag force is present. 
 In our experiments, a projectile impacts a quasi-two-dimensional bed of bidisperse photoelastic grains.  
 We use high-speed imaging to measure high- resolution position data of the projectile trajectory and simultaneously visualize particle-scale force propagation in the granular medium.  
 When the impact angle becomes important, the spatial structure of the stress response reveals relatively weak force chain propagation in the horizontal direction. 
 Based on these observations, we describe the decrease of the inertial drag force with impact angle.

\end{abstract}

\section{Introduction}
Impact onto dry granular materials presents many complexities that are not well-understood, from the disordered contact networks that transmit forces among grains to the flow behavior that can readily change between solid-like rigidity and fluid-like flow \cite{KatBk, vandermeer}. 
For instance, as a solid intruder impacts a dense granular medium, the target exerts a stopping force on the impeding projectile as grains displace and flow. 
The impact of a solid object on granular matter is ubiquitous in nature from locomotion on sand \cite{CLi} to astrophysical crater formation \cite{daniels} and accordingly is being actively explored in experiments  and simulations \cite{KatBk,brzinski,tsimring,zheng2018_pre, wada2006numerical, goldman}.

Such investigations stem from the interest to understand the drag force on an intruder moving through granular media.
The laws of Poncelet have sustained with time as an empirical description that successfully captures governing dynamics of low-velocity granular impact \cite{poncelet}.
More recently, Katsuragi and Durian extended to a phenomenological law to describe forces of normal impacts \cite{KatDur}:
\begin{equation}
F = mg - f(z) - h(z)\dot{z}^2
\end{equation}
The terms of this force law include $mg$ as the force of gravity, friction force $f(z)$ as a function of its depth where $z$ is the depth relative to the top of the granular surface, and a inertial drag force as a function of the square of the intruder velocity and $h(z)$ which represents collisional stress \cite{KatDur,ClarkPRL,Bester}.
The inertial drag force dominates the bulk of deceleration and models momentum transfer through collisions between the projectile to grains.

Recent work examines the connection between local granular response and the dynamics of the projectile via high-speed photoelastic visualization \cite{ClarkPRL,zheng2018_pre}. 
As proposed through experiments by Clark et. al. \cite{clark2014} and Bester et. al. \cite{Bester}, the collisional model relates the inertial drag to the energy dissipated by intermittent collisions with force-carrying clusters of grains during penetration.
Loss of energy to the medium has mean behavior consistent with the empirical force law and connected to the acoustic activity beneath the intruder \cite{ClarkPRL}.
The grain-scale origins of the inertial drag term of the impact force law was then linked to this mechanism \cite{clark2014,Bester}.

The aforementioned work focused on perpendicularly impacting projectiles.
Meanwhile studies of non-normal impacts have been limited, focusing primarily with connections to aeolian sand transport \cite{valance2003}, where the size of the projectile is about the same as the size of target grains.
Prior numerical simulations showed that the empirical drag force model can be extended the horizontal and vertical direction of oblique impact if the impact angle with the granular bed is above 30 degrees with the horizontal \cite{wang2012scaling,wang2015ec}; 
a decomposed model described the resistance force exerted on the projectile impacting at different angles.
Additionally, the effect of the projectile's rotation was investigated using two-dimensional simulation and shown to affect projectile displacement, with penetration depth being greatest when the rotation is zero and the horizontal displacement depending on the direction of rotation \cite{YeWangZhangPRE86(2012)}.

Here we provide the first experimental study of the application of the collisional model to oblique impact using photoelastic imaging.
We examine the kinematics and dynamics of a solid disk impacting obliquely into a two-dimensional photoelastic granular medium. 
The influence of impact velocity and impact angle on the projectile's trajectory is discussed.
We also explore the rarely-studied role of rotation on the penetration of a projectile through granular media.
The data is used to demonstrate the extenstion of the collisional model to the drag force of a projectile impacting granular media at an angle above 40 degrees.

\section{Experiment}
\label{expt}

Figure 1(a) shows a schematic of our experimental setup; the apparatus is similar to previous experiments \cite{ClarkPRL,clarkepl,zheng2018_pre}.
A bronze disk of radius $R$ = 6.3cm impacts a quasi-two-dimensional system of bidisperse photoelastic grains (grain diameters $d$ are 4.3mm and 6mm with a 3mm thickness) which are confined between two Plexiglas sheets separated by a 3.3mm gap. 
The grains are cut from urethane sheets (Vishay Precision group PS-1 with an elastic modulus of 2.5 GPa).
The container width and depth are 1.2m and 0.75m respectively, which is large enough to minimize boundary effects \cite{durian,seguin}.

For each experimental run, a disk travels down a chute that is lined with teflon and strikes the center of the free surface of the granular bed.
A straight edge is used to level the surface before each run.
The impact is recorded with a Photron Fastcam SA5 high-speed camera at 30000 frames per second and with a resolution of 448 x 504 pixels (1 pix = 0.67 mm).
The impact angle of the projectile's trajectory $\theta_{imp}$ is determined by finding the angle formed by the displacement of its center of mass with the horizontal and is varied by shifting the angle of the chute with the granular surface; $\theta_{imp}$ varies from 40 degrees to 90 degrees.

\begin{figure}[!t]
\centering 
 \includegraphics[width=0.46\textwidth]{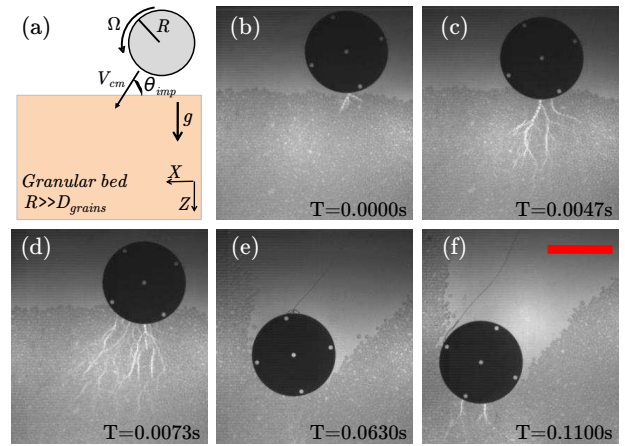}
\caption{(a) Schematic of the experiment. A bronze disk of radius R is released down a quasi-two-dimensional chute placed at $\theta_{imp}$ to a bed of photoelastic bidisperse granular media. It is released from different heights to achieve varying translational velocity $V_{cm,i}$ and rotational speed $\Omega_i$. (b-f) Sequential images showing force network visualization due to intruder penetration. Force chains are intermittently generated along the direction of the projectile as it displaces the medium. Time $T$ is the time after impact. The red scale bar shows 5 cm.}
\label{fig:1}      
\end{figure}

To examine the disk's trajectory, we locate and track both the projectile center of mass position, which is decomposed into horizontal ($\hat{x}$) and vertical ($\hat{z}$) components, and its angular position about the center of mass ($\theta_{proj}$) at each frame.
We measure $\theta_{proj}$ with respect to the horizontal axis and set $\theta_{proj}=0$ at the moment of impact, and the displacement of the projectile is determined relative to its location upon initial contact with the grains.
Velocity $V$ is then calculated by numerical differentiation of high-resolution displacement data.
The total velocity of the disk in a fixed reference frame is $V = V_{cm} + r \times \Omega$, corresponding to the motion of the center of the disk plus its motion about the disk center.
We determine the translational velocity ($V_{cm}$ and $\Omega$) as $V_{cm} = \sqrt{V_x^2 + V_z^2}$, where $V_x$ and $V_z$ are determined from the derivative of the raw position data in the $\hat{x}$ and $\hat{z}$ directions and measurement noise in the data is minimized using a gaussian filter with a width of 2.5\% of the position data; angular velocity $\Omega$ is similarly calculated from $\theta_{proj}$.
The disk is released at different heights above the granular bed to achieve an impact speed $V_{cm,i}$ ranging from 2.5 m/s to 5 m/s.
It also has an initial angular speed $\Omega_i$ upon impact and rotates as it displaces grains.

\begin{figure}
  \includegraphics[width=0.5\textwidth]{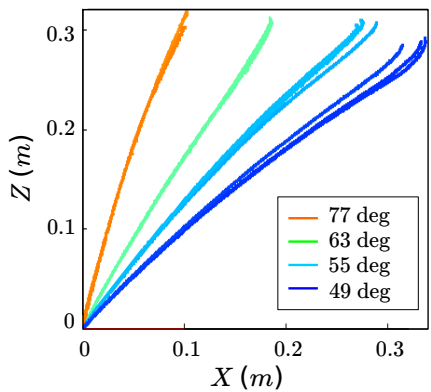}
 \centering
\caption{Trajectories through granular media of a disk (radius R = 6.3cm) at impact speed $V_{cm,i}=4.5\pm 0.2 m/s$. The color of each curve corresponds with $\theta_{imp}$ of the run. The location of initial impact is given as (0,0).}
\label{fig:2}       
\end{figure}

Photoelastic images are used to visualize the granular response to impact.
When polarized light passes through birefringent granular material, the patterns of light indicate the stress state due to the intruder \cite{wood2011,daniels2017}.
Impact reveals the force chain network in which forces are carried along preferred directions of contacting grains.
Figures 1(b) - 1(f) show the evolution of force chain structure for an impact run for which $V_{cm,i}=3$ m/s and $\theta_{imp}=45 ^{\circ}$.
From photoelastic images, we can track the contact region of the intruder during its penetration and show that the force is highly fluctuating as granular networks are excited along the disk \cite{ClarkPRL}. 
Collisions with clusters of force-carrying grains dissipate kinetic energy and stop the intruder.
These force chains span along the direction of the intruder's trajectory and propagates normal to its surface.

\section{Projectile trajectories}
\label{sec:3}
Through a series of experimental runs, we analyze the relationship between $\theta_{imp}$ and trajectories from which we can assess the energy dissipation of the projectile.
In fig. 2, we display Z versus X for typical trajectories for which the disk impacts with similar $V_{cm,i} \approx$ 4.5 m/s and at different $\theta_{imp}$.
Here $Z$ =$X$= 0 m corresponds with the position of the disk center at the moment of initial impact as determined from the photoelastic response.
For each run, the disk follows a linear trajectory, maintaining $\theta_{imp}$ with the horizontal as it comes to rest.
Though it rotates during its motion through the granular medium, this does not influence the trajectory direction.
No net resistance force acts to change the direction of the trajectory during its penetration.
We note that for high $V_{cm,i}$ and low $\theta_{imp}$, trajectories have a minor circuitry at the end of the penetration, indicating slight rebound \cite{seguin}.
As the disk comes to rest, displaced grains then fall onto it; this may influence the rebound.

\begin{figure}[!b]
  \includegraphics[width=0.46\textwidth]{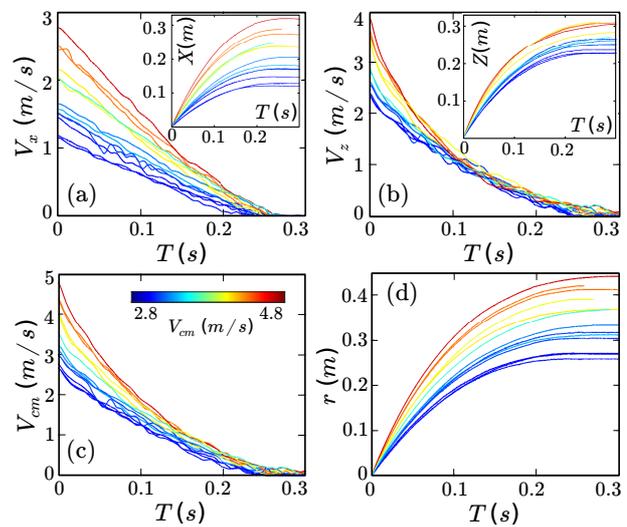}
  \centering
\caption{Velocities of trajectories, where (a) $V_x$ and  (b) $V_z$ are found from derivatives of $X(T)$ and $Z(T)$, respectively and (c) $V_{cm}=\sqrt{V_x^2 + V_Z^2}$ for a projectile with a similar impact angle of 55-60 degrees.
(d) Total displacement $r(T)$, where $r = \sqrt{X^2 + Z^2}$.
Time $T = 0$ is the time of first contact, as determined from the first observation of a photoelastic response. 
Color of each curve in all plots indicates initial translational impact speed, $V_{cm,i}$.}
\label{fig:3}       
\end{figure}

We determine the depth and horizontal displacement from the point of initial contact, as well as the corresponding velocities $V_x$ and $V_z$ as a function of time $T$, with $T$=0 being the moment of impact.
Figure 3 shows the trajectories of a disk impacting with $V_{cm,i}$ varying from 2.8 m/s to 4.8 m/s and $\theta_{imp}$=55 degrees. 
Larger horizontal and vertical displacement occur with increasing $V_{cm,i}$, but the stopping time, the time at which $V_{cm}$ become 0, remains about the same, as described in fig. 5.
We also express total displacement $r$ versus $T$, where $r = \sqrt{X^2 + Z^2}$, in fig. 3(d) and find that $r$ rapidly approaches its maximum penetration within 0.3 seconds.
The total displacement of the disk increases with $V_{cm,i}$ for a chosen $\theta_{imp}$.

The disk rotates as it travels through the grains.
We start by measuring the angular displacement $\theta_{proj}$ of the disk for different impact speeds and at similar $\theta_{imp}$ (see fig. 4(a)).
Positive values correspond with counterclockwise rotation of the projectile.
Maximum rotational displacement tends to be proportional to $V_{cm,i}$, though occasionally low impact speed runs lead to comparatively high angular displacement.
This likely depends on $\Omega_i$ imparted to the disk when propelling it towards the granular bed.
During the trajectory, $\Omega$(T) surprisingly has an approximately linear decline. 
The slope of $\Omega$(T) gives the average angular acceleration $\alpha_{avg}$ for each run.
Correspondingly, we find that the magnitude of $\alpha_{avg}$ tends to linearly increase with increasing $\Omega_{i}$ and $V_{cm,i}$, as shown in fig. 4(b).
This leads to a similar stopping time for all impact runs regardless of $\Omega_{i}$ or $V_{cm,i}$.

\begin{figure}[!]
  \includegraphics[width=0.46\textwidth]{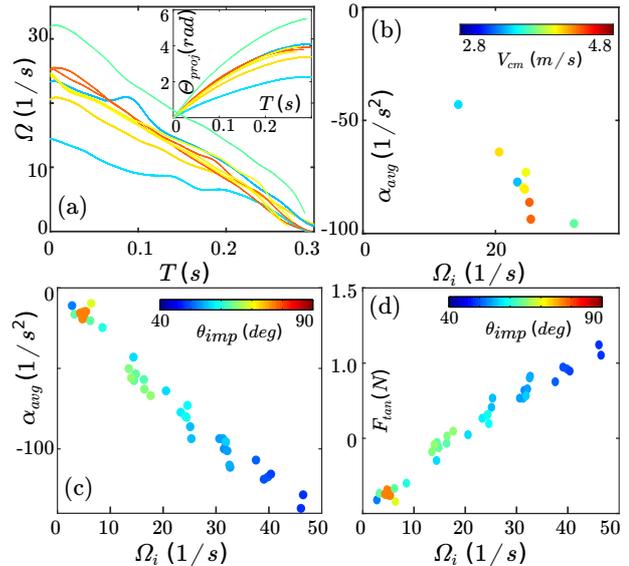}
\caption{(a) Angular speed $\Omega$ versus $T$ for initial impact trajectory angle $\theta_{imp}$ = 55-60 deg. The color of each curve corresponds with the translational impact speed $V_{cm,i}$.
Inset:  Projectile angular displacement, $\theta_{proj}(T)$ of each run.
(b) Average angular acceleration $\alpha_{avg}$ versus angular speed at impact $\Omega_i$, where $\alpha_{avg}$ is calculated from the slope of each $\Omega$(T) curve.
(c) $\alpha_{avg}$ versus $\Omega_i$ at all explored impact angles ($\theta_{imp}$ = 40-90 deg) and speeds (2.5 - 5.0 m/s). The linear relationship is consistent for all $\theta_{imp}$. (d) $F_{tan}$ = I$\alpha$/R versus $\Omega_i$ showing the tangential force magnitude. Figures 4(c) and 4(d) share the same colorbar indicating $\theta_{imp}$.
}
\label{fig:4}       
\end{figure}

From our measurements of the rotational kinematics, we extend to analyze the dynamics of this motion.
A tangential force $F_{tan}$ from the grains acts to accelerate the rotation of the projectile.
Starting with $\alpha_{avg}$, we find that the linear relationship between $\alpha_{avg}$ and $\Omega_i$ persists for all $\theta_{imp}$, as displayed in fig. 4(c).
We then determine $F_{tan}$ from $\alpha_{avg}$ as $F_{tan} = I \alpha_{avg} /R$ where $I$ is the rotational inertia of the disk; this is shown in fig. 4(d).
When comparing $F_{tan}$ to the disk weight, $F_g$ = 2.6 N, its value is at most $31\%$ of $F_g$, increasing with decreasing $\theta_{imp}$ and increasing $\Omega_i$.
This shows the increasing significance of the tangential force with angled impacts.

In previous simulations, the rotation of a projectile was found to influence the dynamic response of the granular bed and increase the projectile's displacement after impact \cite{YeWangZhangPRE86(2012)}.
We accordingly compare to our results to determine if we have similar findings.
We plot the final displacement, $r_{stop}$ versus kinetic energy at impact $K_i$ for varying $\theta_{imp}$ in fig. 5(a).
The maximum penetration, $r_{stop}$, is defined as the distance from the impact point of the disk to its position when $V_{cm}$ becomes 0.
The data follows logarithmic behavior at a particular impact angle, similar to vertical impact experiments \cite{clarkepl}.
Increasing $\theta_{imp}$ (decreasing $\Omega_{i}$) leads to decreasing total displacement at the same total impact energy, which supports results of recent simulations \cite{YeWangZhangPRE86(2012)}.
Stopping time $T_{stop}$ remains approximately constant for a given $\theta_{imp}$; it slightly increases with decreasing $\theta_{imp}$.
Note that the disk stops rotating within 10 ms after it stops translating.

\begin{figure}
  \includegraphics[width=0.46\textwidth]{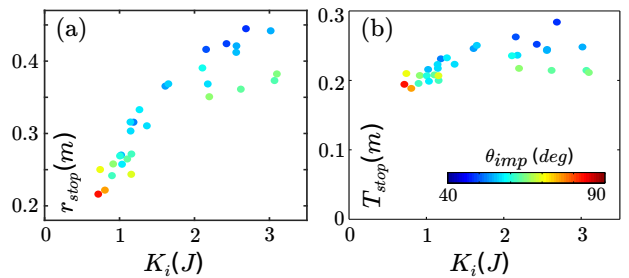}
\caption{(a) Final translational displacement $r_{stop}$ versus initial kinetic energy, with color indicating impact angle. The data are consistent with the logarithmic relationship for a given impact angle; the curve lowers with increasing impact angle. (b) Stopping time $T_{stop}$ at which the translational velocity of the disk reaches zero.}
\label{fig:5}       
\end{figure}

\section{Collisional model of oblique impact}
\label{sec 4}
The application of a collisional-based model to vertical granular impact studies correctly captures the inertial drag force \cite{takehara,clark2014, Bester}.
The foundation of this model is that a projectile transfers momentum to grains by way of sporadic collisions exciting force chains that propagate normal to its local surface.
From our photoelastic imaging results of the granular response to oblique impact, we propose that we can extend the collisional model to different impact angles.
Additionally, since the projectile trajectory entirely follows $\theta_{imp}$, we can treat the net upward force as acting along the $\overrightarrow{r}$ direction to apply the collisional model.

Similar to the work of Clark et. al. \cite{clark2014}, we propose a simple picture to illustrate the drag force acting on the projectile.
In this case, a disk of radius $R$ moves with velocity $V = V_{cm} + r \times \Omega$ at $\theta_{imp}$ with the horizontal through granular media during which it experiences collisions with clusters of grains.
When a collision occurs with a cluster of grains, it acts in the normal direction $\hat{n}$ to the projectile surface.
A collision at the perimeter leads to a change in momentum $\Delta p \propto m_c (V \cdot \hat{n})$, where $m_c$ is the mass of a cluster of grains in time $\Delta t \propto d/(V\cdot \hat{n})$.
A particular collision then has a force given by 

\begin{equation}
f = \frac{\Delta p}{\Delta t} \propto \frac{m_c}{d} (V \cdot \hat{n})^2
\end{equation}
\noindent
Here, $V\cdot \hat{n}$ simplifies to $V_{cm}\cdot \hat{n}$ since the rotational term, $\Omega \cdot (\hat{n} \times \overrightarrow{r})$, is perpendicular to $\hat{n}$.
Since collisions are likely to occur across the projectile surface during its penetration, the total force is given by
\begin{equation}
dF \propto \frac{m_c}{d^2} (V_{cm}\cdot \hat{n})^2 dl \hat{n}
\end{equation}
where $dl$ is the length of the interacting surface.
For a disk, $dl \propto R$ and $F$ has the form $RV_{cm}^2$.
Additionally the length of the interacting surface decreases at low $\theta_{imp}$, as shown, for example, in the photoelastic images of fig. 1.
As grains are displaced for oblique impacts, we observe that collisions are less likely with grain clusters at near horizontal.

\begin{figure}
  \includegraphics[width=0.46\textwidth]{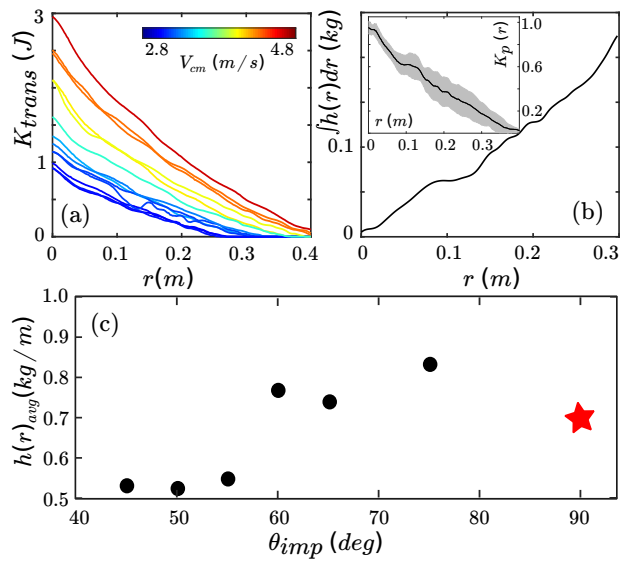}
\caption{(a) Kinetic energy $K$ versus displacement $r$, where $r=0$ at the point of impact. (b) Inset: $K_p$ versus $r$, where $K_p$ is computed for all trajectories as $\Delta K(r)$/$\Delta K_i$ with $\theta_{imp}$ = 50 - 55 degrees. The shaded region shows the standard deviation from the mean (black curve). Main: plot of $\int h(r) dr$ versus $r$ found from $-m/2 log K_p$. The curve is approximately linear. The slope gives $h_{avg}$. (c) $h_{avg}$ versus $\theta_{imp}$ at 5 degree increments.}
\label{fig:6}       
\end{figure}

We connect this picture to the inertial drag coefficient $h$ by fitting $r(T)$ and $V_{cm}(T)$ for all trajectories to the drag force law. 
An approach was recently demonstrated to reformulate the force law as a differential equation of kinetic energy $K$ with respect to displacement to avoid analysis based on the large fluctuations in acceleration data due to intermittent transmissions of acoustic energy along force chains \cite{ambroso,ClarkPRL}.
Accordingly, the force law can modified as follows

\begin{equation}
\frac{dK}{dr} = mg - f(r) - \frac{2h(r)}{m}K
\end{equation}
\noindent
Here we write the equation in terms of total displacement $r$ of oblique impacts.
At each $\theta_{imp}$, the collisional term $h(r)$ is then determined from $K(r)$ trajectories.
We average the difference over all trajectories $K_p$ and obtain $\int h(r) dr$, as  

\begin{equation}
\int h(r) dr = - \frac{m}{2}log K_p
\end{equation}
\noindent
Figure 6(a) shows typical $K(r)$ trajectories with similar $\theta_{imp}$.
By averaging over all pairs of trajectories, we obtain $K_p(r)$ and thus $\int h(r) dr$ at a particular $\theta_{imp}$ using eq. (5), as shown in fig. 6(b).
The slope of the curve gives the average collisional term $h_{avg}$.
We thereby obtain a prediction of inertial drag force for a specific $\theta_{imp}$.
When extending to different $\theta_{imp}$, we similarly calculate $h_{avg}$;
figure 6(c) shows the dependence of $h_{avg}$ on $\theta_{imp}$.
At high $\theta_{imp}$, $h_{avg}$ has an approximately constant value.
There is then a slight decrease of $h_{avg}$ once $\theta_{imp}$ is below 60 degrees, possibly as a smaller length of the disk collides with the granular target.
The decline of $h_{avg}$ with decreasing $\theta_{imp}$ is also supported by our observations of larger $r_{stop}$ and $T_{stop}$ within the $\theta_{imp}$ range.

\section{Conclusion}
\label{sec 5}
We investigate the dynamics of a circular projectile obliquely impacting a two-dimensional photoelastic granular bed. 
Much of its deceleration is dominated by a velocity-dependent inertial drag force, which can be derived from a collisional-based model.
Drag from the granular target is due to sporadic, normal collisions with force-carrying clusters of grains.
This grain-scale mechanism persists as we vary the impact angle of the projectile. 
We extend the collisional model to oblique impacts and determined the influence of impact angle on the collisional term using our experimental data of trajectories at varying impact speed.
As impact angle is decreased from vertical, the inertial drag force slightly decreases, as a shorter length of the projectile perimeter interacts with the granular target.
The results broaden our understanding of the dynamic force response and further supports the collisional model as a connection to the physical origin of the inertial term.
Further experiments should investigate how altering the shape of the projectile changes the trajectory and the force chain patterns. 
This would allow modifications to be made to the collisional model. 

\begin{acknowledgements}
This work is dedicated to Prof. Robert Behringer, whom we are deeply indebted to and will forever miss. His role in supporting and mentoring this research clearly justifies inclusion as a coauthor. This work was funded by NSF Grant No. DMR1206351 and DMR1809762, ARO No. W911NF-18-1-0184, NASA Grant No. NNX15AD38G, the William M. Keck Foundation, and a Duke University Provost's Postdoctoral fellowship (CSB).
\end{acknowledgements}

\bibliographystyle{unsrt}       
\bibliography{2DImpact}   

\begin{thebibliography}{10}

\bibitem{KatBk}
Hiroaki Katsuragi.
\newblock {\em Physics of Soft Impact and Cratering}.
\newblock Springer, Tokyo, 2016.

\bibitem{vandermeer}
Devaraj Van~Der Meer.
\newblock Impact on granular beds.
\newblock {\em Annual Review of Fluid Mechanics}, 49:463--484, 2017.

\bibitem{CLi}
Chen Li, Tingnan Zhang, and Daniel Goldman.
\newblock A terradynamics of legged locomotion on granular media.
\newblock {\em Science}, 339:1408--1411, 2013.

\bibitem{daniels}
Karen~E Daniels, Joyce~E Coppock, and Robert~P Behringer.
\newblock Dynamics of meteor impacts.
\newblock {\em Chaos: An Interdisciplinary Journal of Nonlinear Science},
  14(4):S4--S4, 2004.

\bibitem{brzinski}
T.~Brzinski, P.~Mayor, and D.J. Durian.
\newblock Depth-dependent resistance of granular media to vertical penetration.
\newblock {\em PRL}, 111(168002):168002, 2013.

\bibitem{tsimring}
L.~Tsimring and D.~Volfson.
\newblock Modeling of impact cratering in granular media.
\newblock {\em Powders and Grains}, 2:1215--1223, 2005.

\bibitem{zheng2018_pre}
Hu~Zheng, Dong Wang, David~Z. Chen, Meimei Wang, and Robert~P. Behringer.
\newblock Intruder friction effects on granular impact dynamics.
\newblock {\em Phys. Rev. E}, 98:032904, Sep 2018.

\bibitem{wada2006numerical}
Koji Wada, Hiroki Senshu, and Takafumi Matsui.
\newblock Numerical simulation of impact cratering on granular material.
\newblock {\em Icarus}, 180(2):528--545, 2006.

\bibitem{goldman}
Daniel Goldman and Paul Umbanhowar.
\newblock Scaling and dynamics of sphere and disk impact into granular media.
\newblock {\em Phys. Rev. E}, 77(021308):021308, 2008.

\bibitem{poncelet}
J.V. Poncelet.
\newblock {\em Cours de Mecanique Industrielle}.
\newblock 1829.

\bibitem{KatDur}
Hiroaki Katsuragi and Douglas Durian.
\newblock Unified force law for granular impact cratering.
\newblock {\em Nat. Phys.}, 3(420):420--422, 2007.

\bibitem{ClarkPRL}
Abram Clark, Lou Kondic, and Robert~P. Behringer.
\newblock Particle scale dynamics in granular impact.
\newblock {\em Phys. Rev. Lett.}, 109(238302):238302, 2012.

\bibitem{Bester}
Cacey~Stevens Bester and Robert~P. Behringer.
\newblock Collisional model of energy dissipation in three-dimensional granular
  impact.
\newblock {\em Phys. Rev. E}, 95:032906, Mar 2017.

\bibitem{clark2014}
Abram~H. Clark, Alec~J. Petersen, and Robert~P. Behringer.
\newblock Collisional model for granular impact dynamics.
\newblock {\em Phys. Rev. E}, 89:012201, Jan 2014.

\bibitem{valance2003}
C.~Misbah and A.~Valance.
\newblock Sand ripple dynamics in the case of out-of-equilibrium aeolian
  regimes.
\newblock {\em The European Physical Journal E}, 12(4):523--529, 2003.

\bibitem{wang2012scaling}
Dengming Wang, Xiaoyan Ye, and Xiaojing Zheng.
\newblock The scaling and dynamics of a projectile obliquely impacting a
  granular medium.
\newblock {\em The European Physical Journal E}, 35(1):7, 2012.

\bibitem{wang2015ec}
Xiaoyan Ye, Dengming Wang, and Xiaojing Zheng.
\newblock Effects of density ratio and diameter ratio on penetration of
  rotation projectile obliquely impacting a granular medium.
\newblock {\em Engineering Computations}, 32(4):1025--1040, 2015.

\bibitem{YeWangZhangPRE86(2012)}
Xiaoyan Ye, Dengming Wang, and Xiaojing Zheng.
\newblock Influence of particle rotation on the oblique penetration in granular
  media.
\newblock {\em Physical Review E}, 86(6):061304, 2012.

\bibitem{clarkepl}
Abram~H. Clark and Robert~P. Behringer.
\newblock Granular impact model as an energy-depth relation.
\newblock {\em EPL}, 101(64001):64001, 2013.

\bibitem{durian}
E.~Nelson, H.~Katsuragi, P.~Mayor, and D.~Durian.
\newblock Projectile interactions in granular impact cratering.
\newblock {\em Physical Review Letters}, 101(6):068001, 2008.

\bibitem{seguin}
A.~Seguin, Y.~Bertho, and P.~Gondret.
\newblock Influence of confinement on granular penetration by impact.
\newblock {\em Phys. Rev. E}, 78(010301):010301, 2008.

\bibitem{wood2011}
D.~Lesniewska and D.~Muir Wood.
\newblock Photoelastic and photographic study of a granular material.
\newblock {\em Geotechnique}, 61(7):605--611, 2011.

\bibitem{daniels2017}
Karen Daniels, Jonathan Kollmer, and James Puckett.
\newblock Photoelastic force measurements in granular materials.
\newblock {\em Review of Scientific Instruments}, 88(5):051808, 2017.

\bibitem{takehara}
Y.Takehara, S.~Fujimoto, and K.~Okumura.
\newblock High-velocity drag friction in dense granular media.
\newblock {\em EPL}, 92(44003):44003, 2010.

\bibitem{ambroso}
M.~A. Ambroso, R.~D. Kamien, and D.~J. Durian.
\newblock Dynamics of shallow impact cratering.
\newblock {\em Phys. Rev. E}, 72(041305):041305, 2005.

\end{thebibliography}

\end{document}